\documentclass{article}
\usepackage{arxiv}
\usepackage{textcomp}
\usepackage[utf8]{inputenc} 
\usepackage[T1]{fontenc}    
\usepackage{hyperref}       
\usepackage{url}            
\usepackage{booktabs}       
\usepackage{amsfonts}       
\usepackage{amsmath}
\usepackage{nicefrac}       
\usepackage{microtype}      
\usepackage{lipsum}		
\usepackage{graphicx}
\usepackage{makecell}
\usepackage{float}
\usepackage{cleveref}
\usepackage{geometry}
\usepackage[square,numbers]{natbib}
\usepackage{doi}
\usepackage{titling}
\usepackage[section]{placeins}
\usepackage{xcolor}
\usepackage{needspace}
\creflabelformat{equation}{#2\textup{#1}#3}
\newcommand{\round}[1]{\ensuremath{\lfloor#1\rceil}}

\title{Site-Net: Using global self-attention and real-space supercells to capture long-range interactions in crystal structures}

\author{\textbf{Michael Moran, Michael W. Gaultois,* Vladimir V. Gusev, Matthew J. Rosseinsky} \\
\\
Leverhulme Research Centre for Functional Materials Design \\
Department of Chemistry, University of Liverpool \\
Liverpool, United Kingdom \\
\\
*m.gaultois@liverpool.ac.uk \\}

\renewcommand{\shorttitle}{Site-Net}

\hypersetup{
pdftitle={Site-Net},
pdfsubject={q-bio.NC, q-bio.QM},
pdfauthor={Michael~W.~Gaultois, Michael~Moran},
pdfkeywords={Point-set, Transformer, Self-attention, Crystallography, Machine learning, Deep learning, Property prediction, materials informatics},
}

\newcommand{\beginsupplement}{%
        \setcounter{table}{0}
        \renewcommand{\thetable}{S\arabic{table}}%
        \setcounter{figure}{0}
        \renewcommand{\thefigure}{S\arabic{figure}}%
        \setcounter{equation}{0}
        \setcounter{figure}{0}
        \setcounter{table}{0}
        \setcounter{section}{0}
        \setcounter{page}{1}
     }

\begin{document}
\maketitle

\begin{abstract}
	Site-Net is a transformer architecture that models the periodic crystal structures of inorganic materials as a labelled point set of atoms and relies entirely on global self-attention and geometric information to guide learning. Site-Net processes standard crystallographic information files to generate a large real-space supercell, and the importance of interactions between all atomic sites is flexibly learned by the model for the prediction task presented. The attention mechanism is probed to reveal Site-Net can learn long-range interactions in crystal structures, and that specific attention heads become specialised to deal with primarily short- or long-range interactions. We perform a preliminary hyperparameter search and train Site-Net using a single graphics processing unit (GPU), and show Site-Net achieves state-of-the-art performance on a standard band gap regression task. 
\end{abstract}

\section{Introduction}
\label{sec:intro}
The application of machine learning to materials science has enabled a new paradigm of high throughput property prediction for the screening and identification of new materials. Prediction pipelines based on machine learning models are significantly less computationally intensive than DFT and other physical simulations in much the same way that computational methods can be a faster and cheaper alternative to synthetic investigations \cite{Schleder2019, Choudhary2022}. Consequently, the material discovery process can be significantly accelerated by initial screening and recommendation by machine learning models, which may lead to subsequent validation of promising candidates through physical modelling, and the final demonstration of discovery through preparation in the laboratory \cite{Pollice2021}.

Machine learning models that rely only on the elemental composition have been widely successful and have been applied to a range of property prediction tasks \cite{Wang2021, Jha2018}. The elemental composition of materials is often the most well-characterised feature, and the fixed and limited number of elements mean that the compositions can be readily embedded as a fixed length vector that is amenable to most machine learning methods. However, many properties are strongly dependent on the crystal structure, and composition-based methods do not distinguish between materials with similar or identical compositions yet different crystal structures, such as polymorphs. A classic example is graphite and diamond, which both have a trivial elemental composition of pure carbon but have wildly different physical properties (\textit{e.g.}, band gap, electrical resistivity, thermal conductivity) \cite{kharisov2019carbon}.

The challenge of creating a suitable representation of the crystal structure prevents directly embedding crystal structures for use in property prediction tasks. Specifically, periodic crystal structures have an unbounded number of atoms, and the conventional methods used to describe periodic systems are challenging to represent appropriately for a machine learning algorithm. There is an infinite number of possible unit cells that can be chosen for a given crystal structure, and the varying number of atomic sites between the unit cells of different crystal structures makes it difficult to construct a representation using a fixed length vector. Further, any representation of the unit cell that uses a coordinate system must be invariant to rigid transformations, otherwise simple rotation and translation can lead to different predictions for different descriptions of the same crystal structure \cite{Ropers2021}.

Treating the crystal structure as a graph and using convolution neural networks has shown promising results for predicting properties \cite{Xie2018, Chen2019, Choudhary2021, Louis2020}, and such models now outperform composition-only models where sufficient structural data is available. However, these models rely on an explicitly defined cutoff distance or number of neighbours to define a meaningful interaction between atomic sites in the crystal structure. These graph learning methods were initially applied to molecules, where short-range chemical bonds are often much more significant than long-range interactions \cite{Kearnes2016}. However, extended inorganic solids have many competing interactions at a range of length scales, and many functional properties arise from long-range features of the crystal structure.

In this report, we present Site-Net, a point-set model based on global self-attention augmented with pairwise interactions, where all atomic sites are free to interact with each other. Site-Net uses a physically motivated representation of the crystal structure as a point set of atomic sites, which is separated into ``site features'' containing chemical information (about elements), and ``interaction features'' containing geometric information (about positions). The set of atomic sites is directly ingested without any pre-defined connections, and the importance of interactions between all atomic sites is flexibly learned by the model through global self-attention. The attention mechanism is probed to reveal Site-Net learns long-range interactions, and that specific attention heads become specialized to deal with primarily short- or long-range interactions. This learning leads to state-of-the-art performance, which we assess using the band gap regression task from Matbench \cite{Dunn2020}, where Site-Net achieves a mean absolute error of 0.251\,eV on an 80:20 train:test split of the dataset.  

\section{Methods}
\subsection{Representation construction and featurization}
\label{sec:supercell}

\begin{figure}[!ht]
	\centering
	\includegraphics[width=\textwidth]{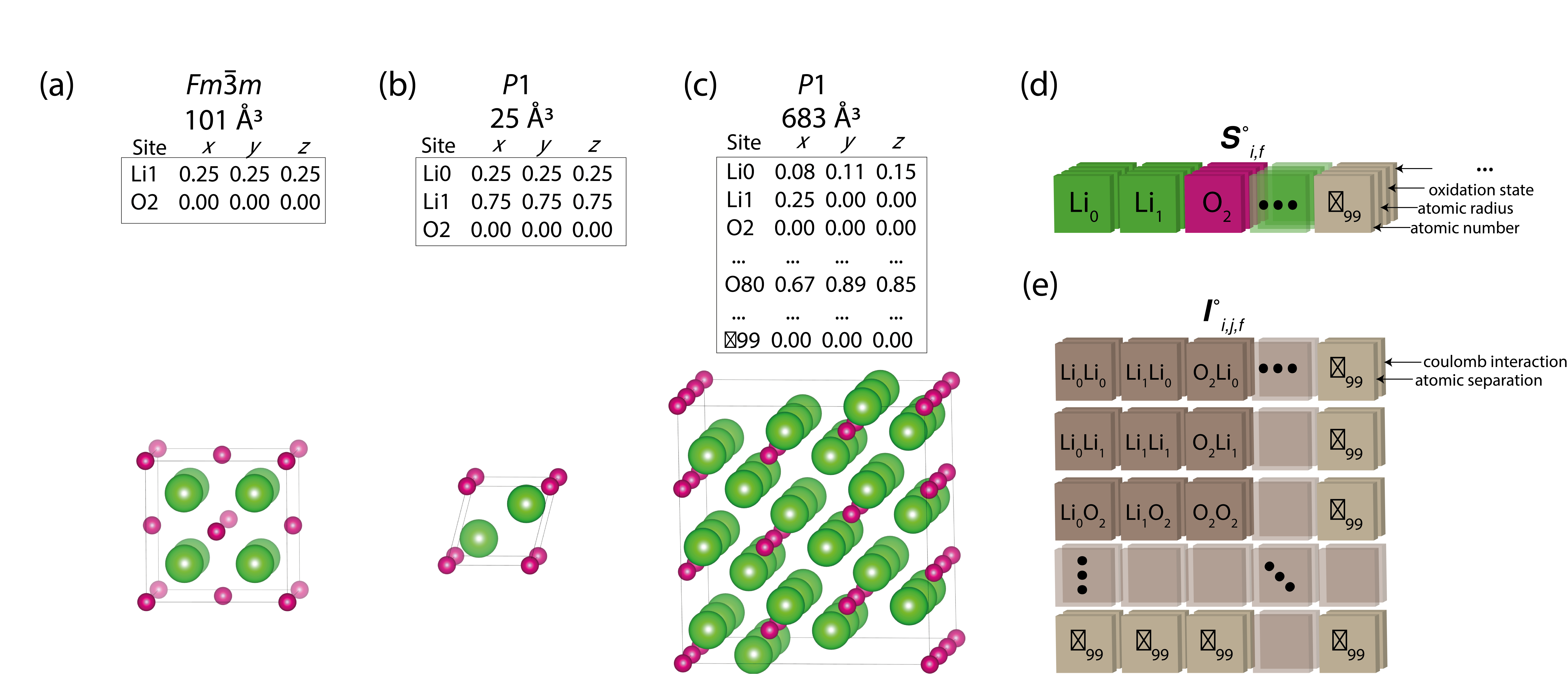}
	\caption[Construction of a suitable representation from the conventional unit cell of a crystal structure.]{Construction of a suitable representation from the conventional unit cell of a crystal structure. (a)  The conventional unit cell of a crystal structure (for example, Li$_2$O) is imported from a crystallographic database. (b) The conventional unit cell is transformed to the primitive unit cell and symmetry is removed. The removal of symmetry to work with raw atoms in space is important to remove any dependence on unit cell choice. (c)  The resulting minimal $P1$ unit cell is expanded to generate a supercell of 100 atoms, which forms the basis of the representation. If exactly 100 atoms cannot be achieved, zero padding is used to bring all atom set to the same size; primitive cells containing more than 100 atoms were not considered during training due to memory constraints. The creation of a supercell is to explicitly include longer range interactions with higher order images. In large set of atoms, the atomic sites and the interactions between these sites are used to construct two tensors respectively. (d) The site features tensor of dimensions [100,101] is denoted by $S^{\circ}_{i,f}$, and (e) the interaction features tensor is denoted by $I^{\circ}_{i,j,f}$, [100,100,2], where $i$ and $j$ represent the site index, and $f$ is the featurization axis. The site features are purely elemental descriptors and do not encode geometry; interaction features enable encoding of the geometric relationship between atoms. The full Euclidean distance matrix (\textit{e.g.}, all pairwise distances) is the foundation of the interaction features and provides a mapping of the sites within the unit cell that is invariant with respect to rigid transformations of any underlying coordinate system.}
	\label{fig:cif2features}
\end{figure}

Periodic crystal structures have an infinite number of atoms and are thus commonly described in a more compact form by choosing an appropriate description (\textit{e.g.}, a unit cell) that can be infinitely tiled in 3 dimensions. There are infinite possible choices of valid unit cells, and although there are several conventions for arriving at a useful unit cell to humans, defining a canonical unit cell for a given crystal structure that is robust to noise is a challenging problem \cite{Ropers2021}. Unfortunately, the lack of a unique unit cell causes issues for training machine learning models, whereas model predictions for a given crystal structure should not be influenced by the arbitrary choice of unit cell. Notably, if the goal is to predict the properties of a crystal structure, two different choices of unit cell for the same crystal structure should lead to the same prediction. With Site-Net, we sidestep this problem by working with a large set of atoms without symmetry, and assume the set is large enough to capture most relevant features without suffering from finite size effects.

Site-Net is able to ingest crystallographic information files (CIF) that are commonly used to represent crystal structures and generate an appropriate representation for training machine learning models (\cref{fig:cif2features}). Any conventional unit cell from a crystallographic database is transformed into a primitive unit cell in $P1$ (\textit{i.e.}, all symmetry constraints are removed, and the atoms are all listed explicitly). This minimal $P1$ unit cell is then iteratively tiled to generate a large set of atoms (\cref{fig:cif2features}c). While Site-Net avoids the need for a canonical choice of unit cell, there is nevertheless a soft requirement to provide each atomic site in the crystal the largest local environment possible. Accordingly, the aforementioned supercell is created to explicitly include longer range interactions with higher order images of the minimal $P1$ unit cell. 

In this work, we show Site-Net performs well with a set of 100 atoms to work within the memory constraints of a single consumer graphics processing unit (GPU), though this is only a technical constraint, and the model performance should improve with increasing number of atoms and the consequently more rich structural context from considering longer range interactions. The set of atoms is generated by determining the optimal transformation of the minimal $P1$ unit cell to the largest possible supercell that is approximately cubic and contains less than 100 atoms.
The creation of appropriate supercells that are roughly cubic remains an open challenge \cite{supercells}, and most methods seek to optimize for a given volume, rather than number of atoms. As this work seeks to optimize for a given number of atoms, we perform supercell construction using an algorithm developed here for this task (\cref{sec:supercell_algo}). If exactly 100 atoms cannot be achieved, zero padding is used to bring all atom sets to the same size. Crystal structures with minimal $P1$ unit cells containing more than 100 atoms were not considered during training due to memory constraints on a single GPU, though these 3245 instances in the Matbench band gap dataset account for only 3\% of the total instances (\cref{fig:size_limit_matbench}). These same large crystal structures were still included during testing and performance evaluation, and thus will penalise the model if it cannot learn about larger structures. 

The resulting set of $\sim$100 atoms, roughly cubic in shape, is featurized into two distinct tensors that separately encode elemental information and spatial information. The elemental information is encoded as a vector of atomic site features (\cref{fig:cif2features}d), consisting of the identities of elements in the crystal structure, along with related properties of these elements. These elemental properties (\textit{e.g.}, atomic radius) can be manually defined, though we also include a learned embedding unique to each element, similar in concept to word embeddings with word2vec \cite{word2vec}, where the tokens are chemical elements. 
For every chemical element, Site-Net stores a unique vector that is updated during model training; the length of the elemental vectors is a hyperparameter of the model (\cref{Tab:hparam}.
In the present implementation of Site-Net, the raw site features are represented by a tensor of dimension [100,101], comprising the number of sites (100), and the elemental features associated with each of the sites (101 in this report).
The spatial information is encoded in the interaction features using a full pairwise interaction matrix between these sites (\cref{fig:cif2features}e). The core of the interaction features is the full real-space Euclidean distance matrix of all atoms (respecting periodic boundary conditions), which ensures the spatial relationships of all atoms in the crystal structure are encoded \cite{7298562, Liberti2014}. 

\begin{figure}[!ht]
	\centering
	\includegraphics[width=\textwidth]{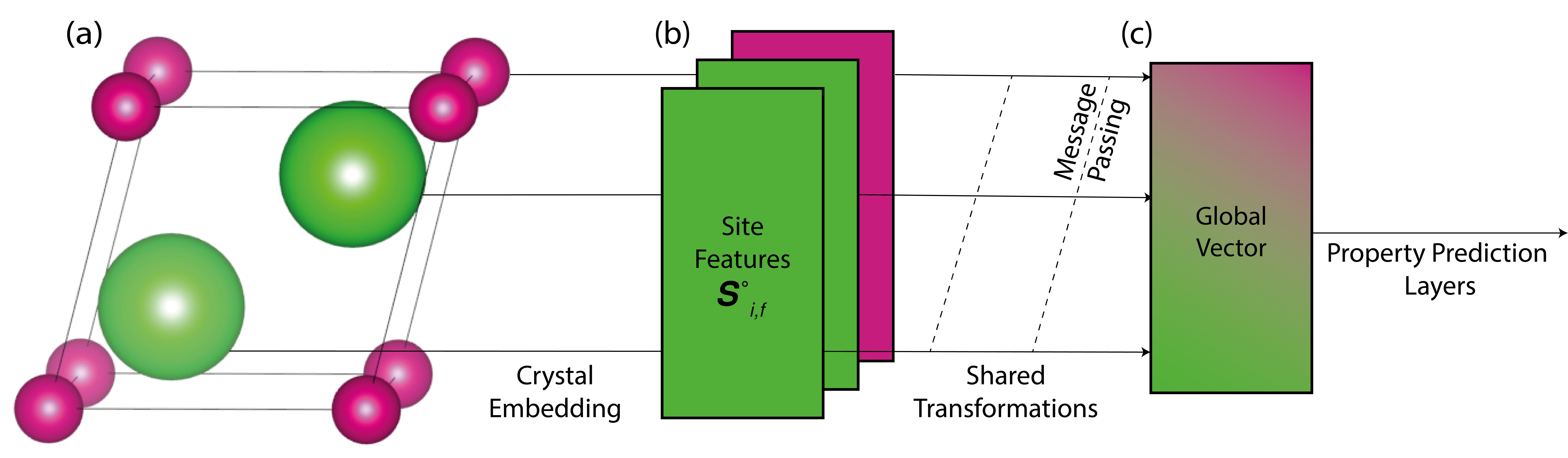}
	\caption[Site-Net transforms a crystal structure into a fixed length global vector that can be used for downstream tasks (such as property prediction).]{Site-Net transforms a crystal structure into a fixed length global vector that can be used for downstream tasks (such as property prediction). This is simplified here for illustrative purposes. (a) The minimal $P1$ unit cell is used to generate site features, shown here for Li$_2$O, which has 2 chemically equivalent Li sites, and one O site. (b) These vectors of site features are then passed through a set of neural networks (including self-attention blocks) to create new context-enriched site features that are imbued with knowledge of their chemical and structural environment. (c) These context-enriched site features are then compressed by a permutation-invariant function (such as taking the mean) to generate a fixed length global vector that describes the entire crystal structure. Without the multiple layers of self-attention to enrich the context of the features, mean pooling of the raw features into a global vector would otherwise cause too much information loss for useful property predictions.}
	\label{fig:model_concept}
\end{figure}
\FloatBarrier
\subsection{Self-attention as a mechanism to create context-enriched site features}
\label{sec:self-attention}
\begin{figure}[!ht]
	\centering
	\includegraphics[width=\textwidth]{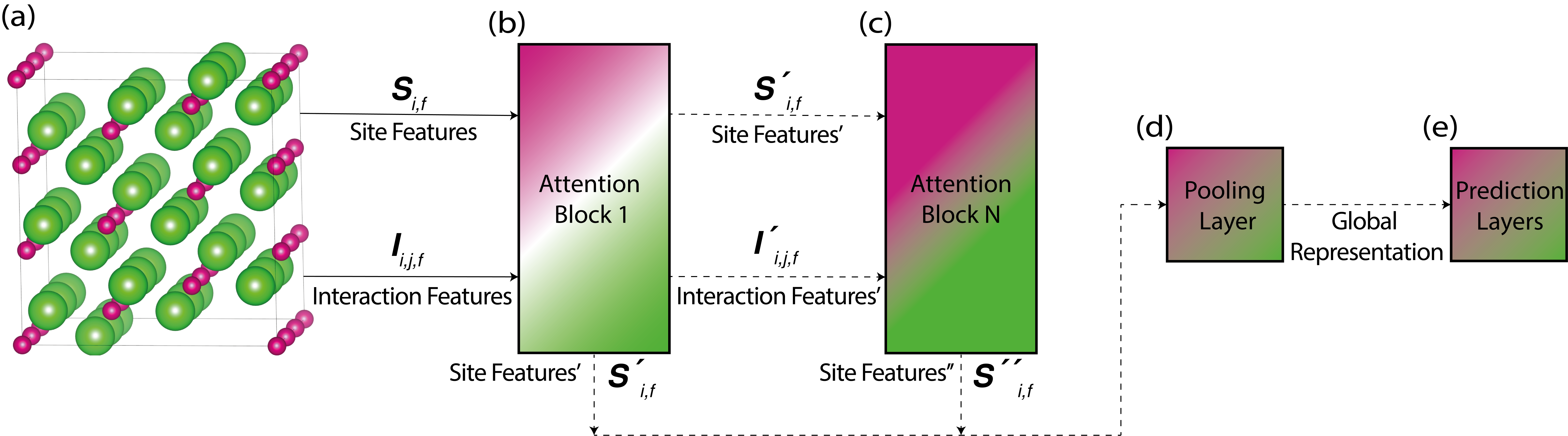}
	\caption[A simplified architecture of Site-Net shows the dataflow from atoms in a crystal structure, to a downstream property prediction task.]{
	A simplified architecture of Site-Net shows the dataflow from atoms in a crystal structure, to a downstream property prediction task. (a) The site features and interaction features are generated from the list of atoms in a large supercell (for example, Li$_2$O). These are passed to attention blocks, which are used to progressively enrich atomic site features with contextual information from neighbouring atomic sites. (b) The first attention block generates new site and interaction features, which are sequentially fed into subsequent attention blocks. After passing through the final attention block, the site feature outputs from all attention blocks are concatenated together. (d) The concatenated site features are then passed to a pooling layer for permutation-invariant aggregation, where the mean is taken to produce a fixed length global feature vector that describes the crystal in its totality. (e) This global feature vector is then processed downstream through prediction neural network layers to generate predictions for a property of interest.
	}
	\label{fig:model_outline}
\end{figure}

Site-Net is a set transformer \cite{NIPS2017_3f5ee243} architecture that takes a crystal structure, constructs a point set from the atomic sites in the unit cell, and processes the point set into a fixed length global feature vector representing the entire crystal structure, which is suitable for downstream property prediction (\cref{fig:model_concept}). Rather than encoding spatial information through a coordinate system (\textit{e.g.}, PointNet \cite{pointnet}), a matrix of pairwise interactions is incorporated into the custom attention mechanism to iteratively encode the spatial information into the point-set.
Once a fixed length global vector has been attained, property prediction is performed through the application of standard dense neural network layers. However, the compression into a fixed length global feature vector necessary for property prediction is performed using permutation-invariant aggregation (here we use mean pooling), which leads to significant information loss. 
For example, taking the mean over the initial site features would destroy most relevant crystallographic information. Accordingly, Site-Net passes the initial site features through multiple attention layers to enrich the atomic site features with the context of their local chemical and structural environment (\cref{fig:model_outline}). These context-enriched site features retain enough structural information following compression to allow useful property predictions.

Before being passed to an attention block, the raw site features ($S^{\circ}_{i,f}$) and raw interaction features ($I^{\circ}_{i,j,f}$) are reprocessed to an auxiliary embedding. Here, $i$ and $j$ are atomic site identities, and $f$ is the featurization dimension. The auxiliary embedding is likely to depend on the prediction task, so the lengths of the featurization dimensions are tunable hyperparameters to allow the Site-Net model flexibility to find an optimal representation or dimensionality for a given task. This is accomplished by a single neural network layer preceding the first attention block, which ingests the raw site features ($S^{\circ}_{i,f}$) and interaction features ($I^{\circ}_{i,j,f}$) and generates processed analogues of the correct dimensionality.
Specifically, the raw site feature tensor $S^{\circ}_{i,f}$ [100, 101] is transformed to $S_{i,f}$ [100, $\lambda$], and the raw interaction features tensor $I^{\circ}_{i,j,f}$ [100, 100, 2] is transformed to $I_{i,j,f}$ [100, 100, $\mu$]. 
These dimensions are consistent across the attention blocks for both input and output. In the final model presented here, the hyperparameters found after a preliminary search are $\lambda$\,=\,90 and $\mu$\,=\,48 (\cref{Tab:hparam}). 

\begin{figure}[!ht]
	\centering
	\includegraphics[width=\textwidth]{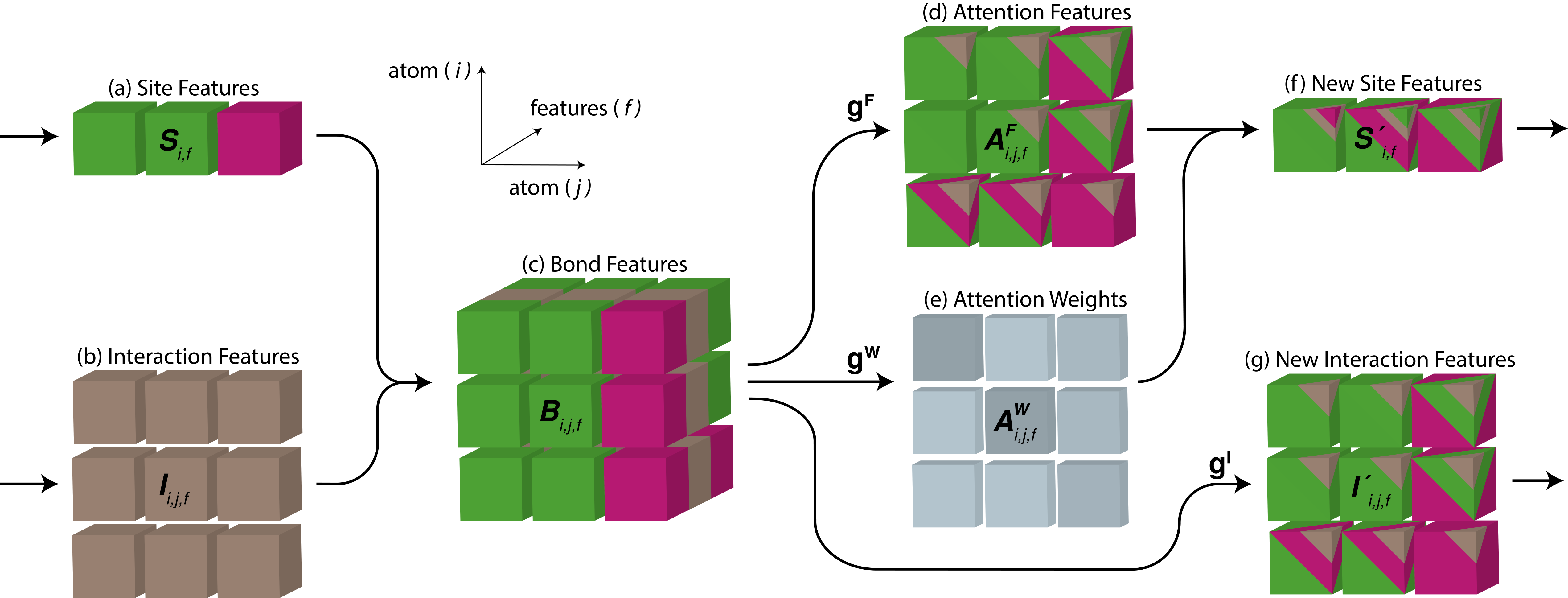}
	\caption[The mechanism of the Site-Net attention block, whose purpose is to enrich the atomic site features with context of their local environments, is illustrated using the simplified example of the minimal $P1$ unit cell of Li$_2$O.]
	{The mechanism of the Site-Net attention block, whose purpose is to enrich the atomic site features with context of their local environments, is illustrated using the simplified example of the minimal $P1$ unit cell of Li$_2$O.
	Each attention block ingests (a) the site features $S_{i,j}$ and (b) the interaction features $I_{i,j,f}$, which are concatenated to generate the bond features. 
	(c) The bond features $B_{i,j,f}$ are a unified representation of every ordered pair of atoms and the interaction between them.
	These set of bond vectors $B_{i,j,*}$ then go through a series of transformations to eventually generate the new site features $S^\prime_{i,j}$ and new interaction features $I^\prime_{i,j,f}$.
	(d) The attention features $A^F_{i,j,f}$ are derived from the learned function $g^{F}$ and serve as precursors to the new site features.
	Self-attention mechanism is used to combine these attention features as a weighted sum to form $S^\prime_{i,j}$. 
	(e) The relative contribution of each attention feature is dictated by the scalar attention weights $A^W_{i,j,f}$. These weights are derived from the learned function $g^W$, and represent the strength of the influence of a particular attention feature on the local environment.
	(f) New site features $S^\prime_{i,j}$ produced through such a self-attention mechanism represent local environments of the atomic sites instead of solely elemental properties.
	(g) The new interaction features $I^\prime_{i,j,f}$ are derived from the learned function $g^I$ to compress the bond features $B_{i,j,f}$ back down to the proper dimensionality, and are enriched by the context of atoms connected by the interactions.
	}
	\label{fig:attention_mechanism}
\end{figure}

\FloatBarrier
Starting from the raw site features in a crystal structure, site features enriched with the context of their local environment are constructed using a sequence of self-attention blocks, where the site features are iteratively replaced with a weighted aggregation of the pairwise interactions with all other atomic sites in the crystal structure representation (\cref{fig:attention_mechanism}). This process replaces the purely elemental features of atomic sites with the aggregation of their local environment, and thus encodes information about the crystal structure into the context-enriched site features. This aggregation function does not depend on the ordering of atomic sites, and is thus permutation invariant in the same way the global feature vector is produced from the site features. At a conceptual level, self-attention is a learned permutation- invariant function that prioritizes the most important interactions when constructing the new enriched site features.

\FloatBarrier
\begin{align}\label{eq:S}
S_{i,f} &\in \mathbb{R}^{N \times \lambda} \\
\label{eq:I}
I_{i,j,f} &\in \mathbb{R}^{N \times N \times \mu} \\
\label{eq:B}
B_{i,j,*} = S_{i,*} \mathbin\Vert I_{i,j,*} \mathbin\Vert S_{j,*}&, \text{ where } B_{i,j,f} \in \mathbb{R}^{N \times N \times (2\lambda + \mu)}
\end{align}
\FloatBarrier

To begin, the site features $S_{i,f}$ (\cref{eq:S}) and interaction features $I_{i,j,f}$ (\cref{eq:I}) for each pair of atoms are concatenated to create bond features $B_{i,j,f}$ (\cref{eq:B}).
The bond feature vector $B_{i,j,*}$ captures interactions between atomic sites $i$ and $j$, and is an ordered combination of a site vector $S_{i,*}$, followed by the interaction vector $I_{i,j,*}$, and then $S_{j,*}$. Here, an asterisk ($\ast$) denotes the span of an index.
Importantly, because the order of the atom pairs is preserved, these bond features are directional ($B_{i,j,*} \neq B_{j,i,*}$). 
Assembling all the bond feature vectors into the complete bond features tensor $B_{i,j,f}$ leads to a unified representation of the crystal structure (\cref{fig:attention_mechanism}c). This is carried forward and subsequently used to derive new context-enriched site features $S^\prime_{i,f}$ and new context-enriched interaction features $I^\prime_{i,j,f}$ (\cref{fig:attention_mechanism}).

\begin{align}
\label{eq:Af}
A^F_{i,j,*} = g^F(B_{i,j,*})&, \text{ where } A^F_{i,j,f} \in \mathbb{R}^{N \times N \times \lambda} \\
\label{eq:Aw}
a^W_{i,j} = \frac{e^{g^W(B_{i,j,*})+\delta_{ij}}}{\sum_j e^{g^W(B_{i,j,*})+\delta_{ij}}}&, \text{ where } A^W_{i,j} \in \mathbb{R}^{N \times N} \text{ and } \delta_{ij} \text{ is the Kronecker delta} \\
\label{eq:Sprime}
S^\prime_{i,*} = \sum_j a^W_{i,j}A^F_{i,j,*}&, \text{ where } S^\prime_{i,f} \in \mathbb{R}^{N \times \lambda}
\end{align}

Global self-attention is used to generate new context-enriched site features $S^\prime_{i,j}$. In this implementation, we introduce intermediate attention features $A^F_{i,j,f}$ (\cref{fig:attention_mechanism}d, \cref{eq:Af}), and attention weights $a^W_{i,j}$ (\cref{fig:attention_mechanism}e, \cref{eq:Aw}).
The vectors $A^F_{i,j,*}$ have the same dimension as site feature vectors and are obtained from bond vectors $B_{i,j,*}$ by means of a fully connected neural network $g^F: \mathbb{R}^{2\lambda+\mu} \xrightarrow[]{} \mathbb{R}^{\lambda}$.
The relative importance of site $j$ to $i$ based on their interaction is captured by the scalar attention weights $a^W_{i,j}$ (\cref{fig:attention_mechanism}e, \cref{eq:Aw}), 
which are computed using another fully connected neural network $g^W: \mathbb{R}^{2\lambda+\mu} \xrightarrow[]{} \mathbb{R}$. 
The number of layers and the number of neurons per layer for $g^W$ are hyperparameters of the model.
The resulting scalar values $g^W (B_{i,j,*})$ are normalized using the softmax function (\cref{eq:Aw}). 
For every atomic site $i$, this softmax normalization ensures that the weights $a^W_{i,j}$ over all atomic sites $j$ sum to 1.
As a consequence of the softmax normalization to generate attention weights $a^W_{i,j}$, the resulting distribution of weights is conceptually similar to a probability distribution over all neighbours, where the attention weights $a^W_{i,j}$ represent the significance of neighbour $j$ to $i$.
Critically, the exponential nature of softmax is likely important to discard many of the negligible contributions that will be present when considering all pairwise interactions in Site-Net. 
To improve training, the attention weights for $i=j$ are increased by 1 to bias the model towards preserving the identity.

Finally, the new context-enriched site feature vector $S^\prime_{i,*}$ is a sum of vectors $A^F_{i,j,*}$ weighted by scalars $a^W_{i,j}$ (\cref{eq:Sprime}).
In simple terms, each atomic site has a vector representing its chemical and geometric configuration, which is subsequently replaced by the mean of all vectors for every neighbour and itself. This mean is modified by the relative importance of every site.
As a consequence, the new site features are no longer a descriptor of a single site. Rather, they are representations of every local environment in the crystal structure. With repeated attention blocks, the representation of each individual site feature becomes more abstract.


\begin{equation}\label{eq:IPrime}
I^\prime_{i,j,*} = g^I(B_{i,j,*}), \text{ where } I^\prime_{i,j,f} \in \mathbb{R}^{N \times N \times \mu}
\end{equation}

The bond features are also used to produce new interaction features $I^\prime_{i,j,f}$. In comparison to the bond features, the obtaining of new interaction features is straightforward. The new interaction features and the bond features are of the same dimension, so new interaction features are obtained (\cref{eq:IPrime}) by passing the bond features through a single feed forward layer ($g^I$) so they are of the expected dimensionality. These new interaction features contain the information of the two sites connected by that interaction and serve a similar role to residual connections, as they preserve this information for subsequent attention blocks.

With respect to the overall architecture, we have described the process of performing single-headed attention. This process can be generalized to multi-headed attention, where multiple sets of attention feature and attention weight tensors are independently computed inside the same attention block and then concatenated.
The use of more attention heads allow more attention operations to be performed in parallel, where each head can focus on a specific group of interactions. 
Similarly, the number of attention blocks can be increased to achieve more abstract features, as attention is performed on the outputs of the last attention block. 
The preliminary hyperparameter search performed on the band gap prediction task revealed that 2 attention blocks and 3 attention heads is a reasonable balance able to achieve state-of-the-art performance (\cref{Tab:hparam}).

\subsection{Post-attention processing and pooling}
\label{sec:post-attention_pooling}

\begin{figure}[!ht]
	\centering
	\includegraphics[width=0.62\textwidth]{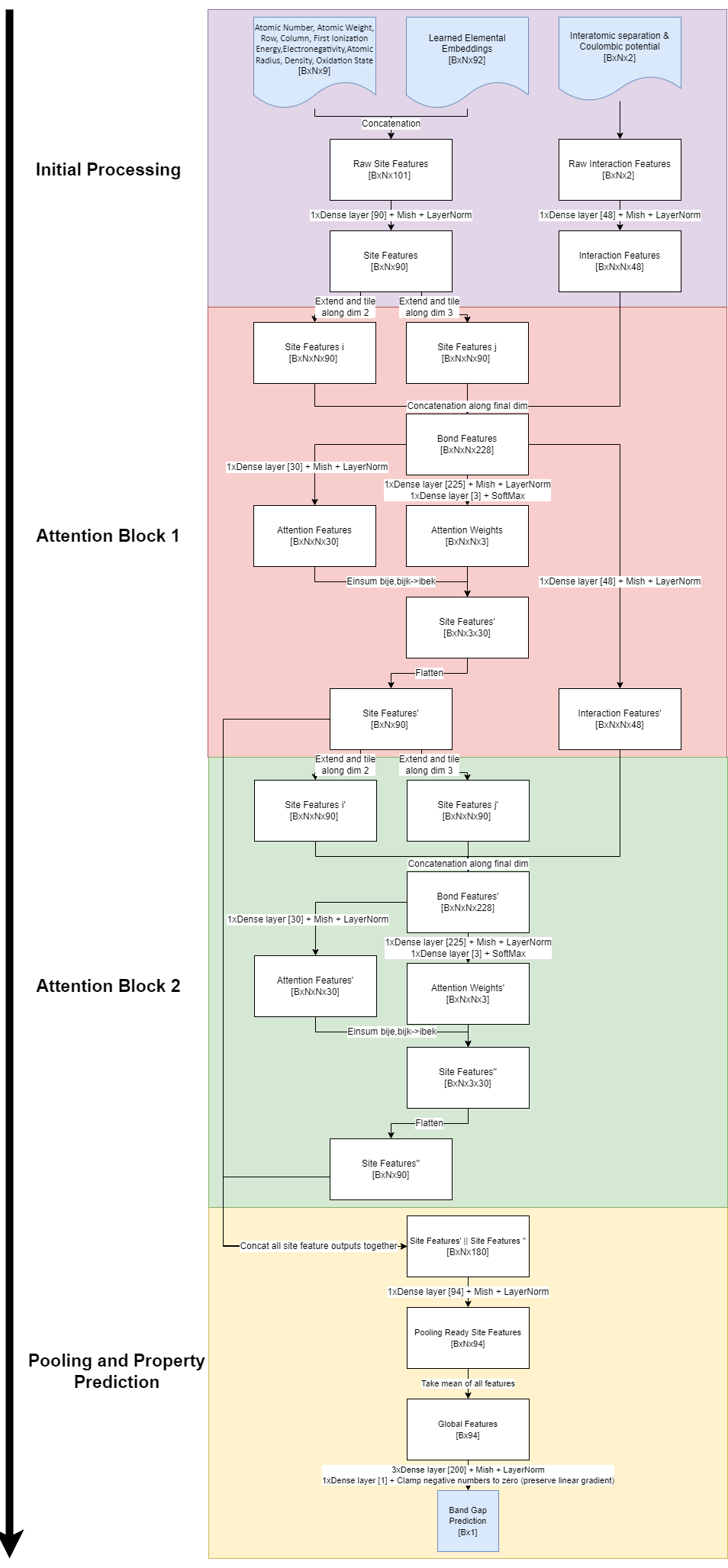}
	\caption[The complete Site-Net model graph for the benchmarking model trained on band gap, with all hyperparameters shown numerically.]{
	The complete Site-Net model graph for the benchmarking model trained on band gap, with all hyperparameters shown numerically. 
	B and N represent the batch size and the number of atomic sites in the unit cell, respectively. These are dictated by the data structure during training, and can be arbitrarily set during inference. For simplicity, only the output dimension size for neural net layers is shown. Dense layers are always applied to the rightmost rank of the tensor.}
	\label{fig:Sflowchart}
\end{figure}

The new site features and interaction features generated by the attention block are fed into the next attention block to repeat this process of contextual enrichment through the construction of higher-level features. After passing through all attention blocks, the separate site feature outputs from each and every attention block are concatenated together in preparation for pooling to a fixed length global feature vector by taking the mean of all sites. Further, a final pre-pooling step is performed to minimize the information loss of the subsequent pooling. 
Here, a simple neural network whose size is a hyperparameter of the model is used to process the concatenated site features from the attention blocks to an auxiliary embedding for pooling, much in the same way that a single neural network layer was used to reprocess the raw site features and interaction features to an auxiliary embedding for performing attention. 
After taking the mean of all sites to produce the fixed length global feature vector, obtaining a property prediction is straightforward a process with a sequence of feed forward neural network layers. An explicit process flow diagram for the entire architecture can be found in the Supporting Information (\cref{fig:Sflowchart}). 

\subsection{Implementation}
\label{sec:implementation}

To generate the representations of the crystals, CIF files are first converted into Pymatgen structure objects \cite{ONG2013314}. From these Pymatgen structure objects, the crystal structure can be featurized using featurization libraries such as matminer \cite{WARD201860} and dscribe \cite{HIMANEN2020106949}. The models were developed using pytorch \cite{NEURIPS2019_9015} combined with the use of the pytorch lightning framework \cite{falcon2019pytorch} to provide automatic GPU training and data management. To handle the varying sizes of crystal structures in the dataset, the tensor representation of atom sets with less than 100 atoms were zero-padded to ensure all tensors were of the same size with respect to the forward pass. All operations performed on padding are excluded from training. 

Hyperparameter tuning was handled via hyperopt \cite{Bergstra_2015} using the Ray Tune \cite{liaw2018tune} distributed hyperparameter tuning framework as a front end. The hyperparameters that performed best on the validation set when trained on the training dataset are benchmarked on the holdout dataset. The hyperparameter search was performed on the Barkla compute cluster using a single Tesla P100 GPU with 16\,GB of VRAM; the batch size was limited by the available VRAM. A preliminary hyperparameter search was performed by sequentially training 30 models for 24 hours each, using previous models to inform future hyperparameter choices. The best set of hyperparameters was then carried forward for longer training of the final models presented here (\cref{Tab:hparam}). The model is sensitive to the choice of hyperparameters, and based on the limited search performed here, these hyperparameters are likely far from optimal and will allow considerable model improvement in the future.

\begin{table}[!ht]
    \caption[Site-Net hyperparameter search space and final values for the reported band gap prediction task.]{Site-Net hyperparameter search space and final values for the reported band gap prediction task. The model was generally sensitive to hyperparameters, which were fixed after achieving best-in-class performance in a preliminary search. All hyperparameters were optimized using Ray Tune \cite{liaw2018tune}, except where noted as fixed. Given the sensitivity of the model to hyperparameter choice and the large search space available, further hyperparameter tuning will undoubtedly improve model performance. 
    } 
    \centering
    \renewcommand{\arraystretch}{1.5}
    \begin{tabular}{p{0.4\textwidth}p{0.3\textwidth}p{0.2\textwidth}}
    \hline
    \hline
        Hyperparameter & value used & range searched \\ \hline
        Site features (From Pymatgen \cite{ONG2013314} \& Matminer \cite{WARD201860}) & 101: Atomic number, Atomic weight, Row, Column, First ionization energy, Electronegativity, Atomic radius, Density, Oxidation state, Learned embedding (92) & Fixed \\ 
        Length of learned embedding & 92 & 1 to 128 dimensions \\ 
        Site features length (attention block) & 90 & 4 to 32 per attention head \\ 
        Interaction features (from Pymatgen \cite{ONG2013314}) & 2: Distance matrix, log(Coulomb matrix) & Fixed \\ 
        Interaction features length (attention block) & 48 & 4 to 32 per attention head \\ 
        Attention blocks & 2 & 1 to 3 blocks \\ 
        Attention heads & 3 & 1 to 8 heads \\ 
        Attention weights network ($g^W$) [depth, width] & [1, 225] & 0 to 3 layers, 32 to 256 neurons per layer \\ 
        Pre-pooling network [depth, width] & [1, 94] & 1 to 4 layers, 32 to 256 neurons per layer \\ 
        Post-pooling network [depth, width] & [3, 200] & 1 to 4 layers, 32 to 256 neurons per layer \\ 
        Activation function & Mish \cite{mish} & Fixed \\ 
        Optimizer & AdamW & Fixed \\ 
        Learning rate & 8.12$\times$10$^{-4}$ & 5$\times$10$^{-5}$ to 10$^{-2}$ \\ 
        Normalization method & LayerNorm \cite{LayerNorm} & BatchNorm \cite{BatchNorm}, LayerNorm\cite{LayerNorm}, None \\ 
        Global Pooling function & Mean & Mean, Max, Self-attention  \\ 
        Batch Size & 18 & 8 to 24 crystal structures \\ \hline \hline
    \end{tabular}
    \label{Tab:hparam}
\end{table}
\section{Results and Discussion}
\label{sec:results_discussion}
The  performance of Site-Net was assessed using the band gap regression task from Matbench, a materials benchmarking test suite \cite{Dunn2020}. The first fold of the preset cross validation pipeline was used for this benchmarking, and consists of 106113 crystal structures and associated band gap energies. The training set was 80\% of the available data and the test set was 20\%. Within the training data, 80\% was used for training, while 20\% of the training data was used for a validation score for hyperparameter optimisation. As training was done using a single GPU, it was computationally unfeasible to run a separate hyperparameter search over all five Matbench data folds. Further, using a single hyperparameter set on all five folds would be unsuitable owing to data leakage, as training data from the first fold---on which hyperparameters are determined---is cycled into test data of the other 4 folds.

The Matbench band gap dataset poses unique challenges as it contains a smooth continuum of positive band gap energies together with a large number of zeros. We employ a custom activation function to address this unique property of the dataset, wherein negative predictions of band gap were clamped to zero while preserving the gradient to allow the model to recover from false zero predictions. Given negative band gaps are non-physical, we thus treat negative predictions as a level of confidence in the classification of zero rather than an ``overshoot'' that needs to be corrected.   

\begin{figure}[!ht]
	\centering
	\includegraphics[width=\textwidth]{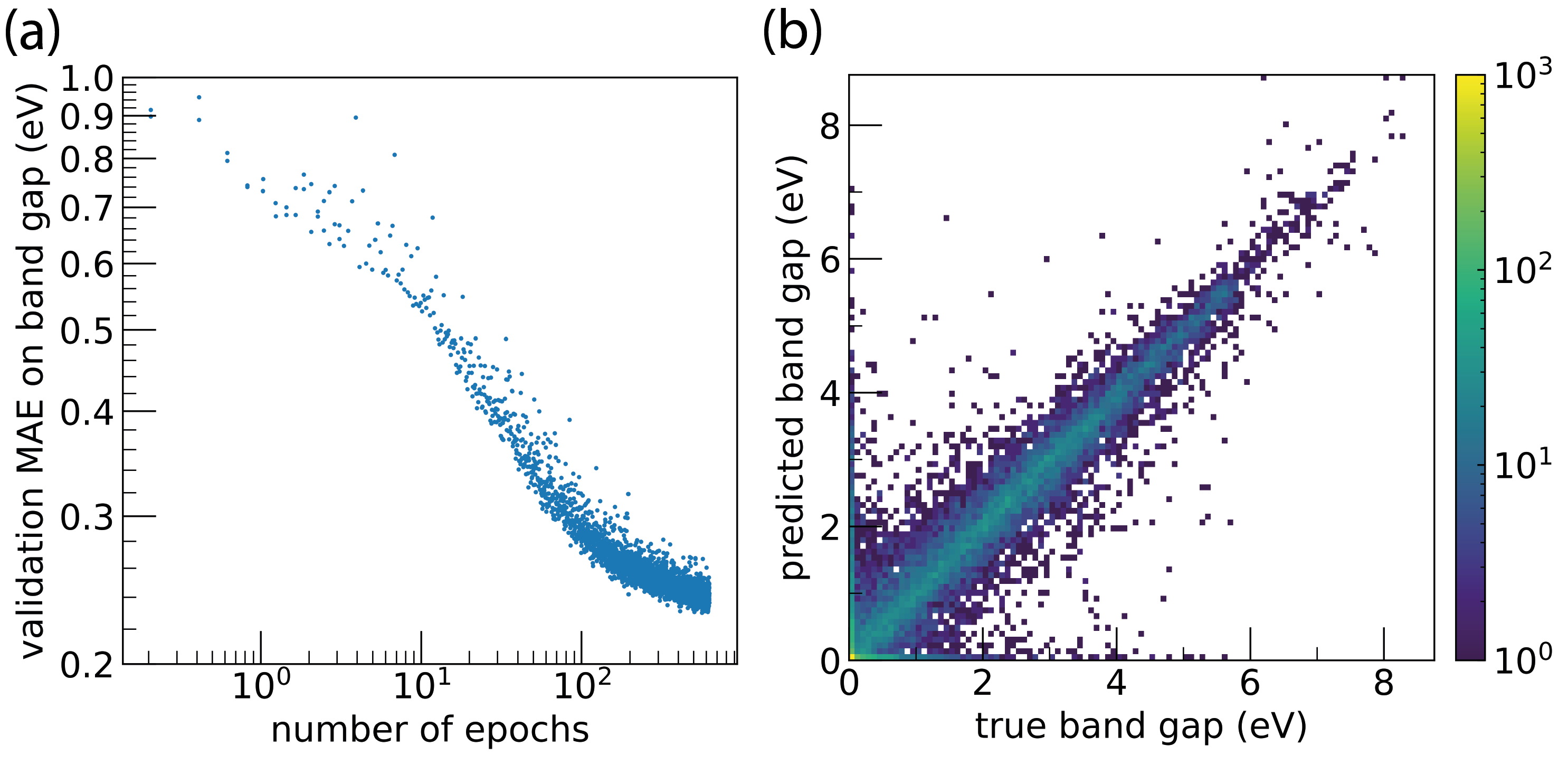}
	\caption[Training and prediction performance of Site-Net on the Matbench band gap prediction task.]{Training and prediction performance of Site-Net on the Matbench band gap prediction task. (a) The learning curve exhibits smooth monotonic loss per epoch, with no overtraining. The mean absolute error (MAE) reaches a plateau at 0.251\,eV after $\sim$500 epochs, which is $\sim$7 days of training. (b) The parity plot reveals the  model is consistent across band gap values, and has an associated MAE of 0.251\,eV. Colour is used to represent the number of materials at a particular coordinate; the peak at the origin is outside the bounds of the scaling used due to the high number of materials in the dataset with a band gap of exactly zero. }
	\label{fig:performance}
\end{figure}

Training Site-Net on the band gap regression task leads to a smooth, monotonic learning curve that steadily converges to a plateau; models did not exhibit divergent overtraining behaviour (\cref{fig:performance}a). Despite its complexity, the model trains to a stable state and does not suffer from problems typically encountered with continued training where the validation score begins to diverge. Site-Net achieves a mean absolute error (MAE) of 0.251\,eV on the band gap regression task, and performance of the model is consistent across band gap values (\cref{fig:performance}b). Even with only a preliminary hyperparameter search, Site-Net currently demonstrates competitive performance with the highest performing algorithms on the leaderboard. For example, CGCNN \cite{cgcnn} has a reported MAE of 0.297\,eV as of this report, and ALIGNN \cite{Choudhary2021}, which uses second and third order interactions such as angles and solid angles between atoms, is the highest performing algorithm with a reported MAE of 0.186\,eV.

\begin{figure}[!ht]
	\centering
	\includegraphics[width=\textwidth]{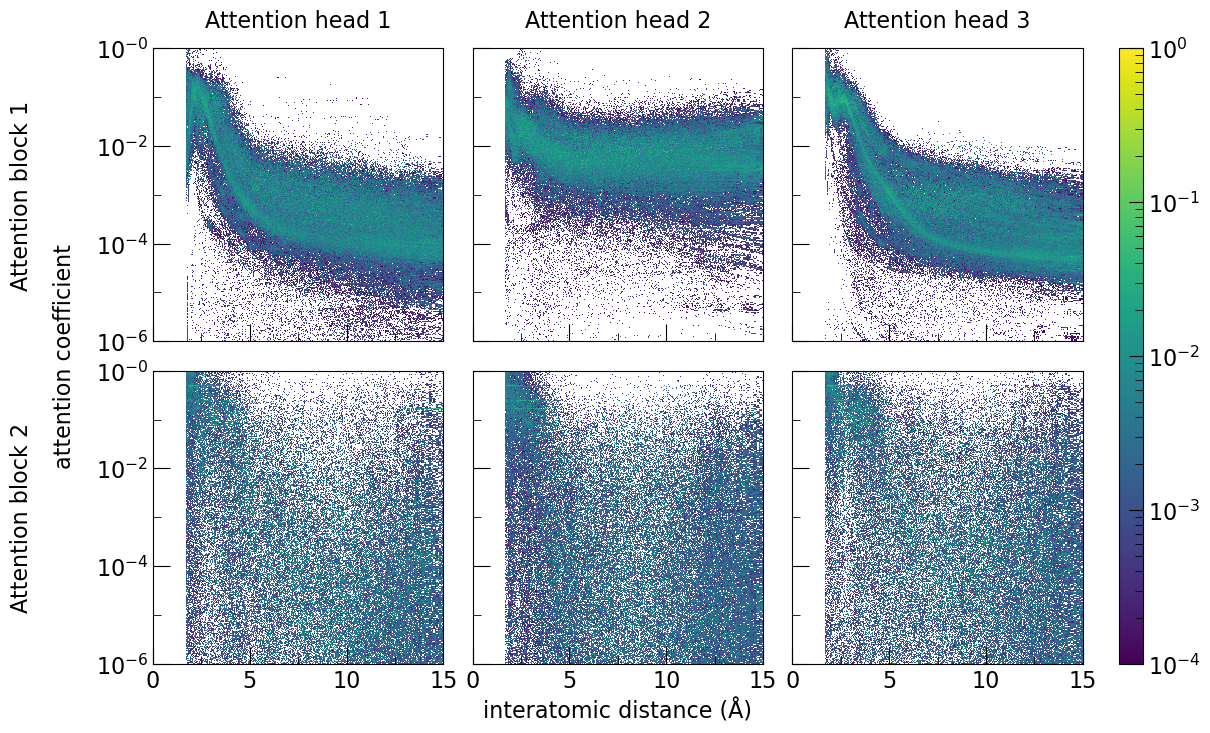}
	\caption[The attention weights of the trained band gap model for all pairs of atomic sites in the test dataset as a function of inter-atomic distance.]{
	The attention weights of the trained band gap model for all pairs of atomic sites in the test dataset as a function of inter-atomic distance. 
	To visualize the $\sim$10$^7$ attention weights, 2-dimensional histograms are constructed by ordinally binning by the interatomic distance, and then normalizing such that the sum of all attention weights at any one distance bin sum to 1.
	The colour corresponds to the proportion of attention weights that lie within a given bin.
	The number of attention heads (3 in this report) and attention blocks (2 in this report) are hyperparameters of the model and were chosen by a preliminary hyperparameter search. The attention heads of the first attention block focus on atomic sites that are close together ($<$5\,\AA{}), which is consistent with local interactions being important to material properties. The second attention head notably contains more long-range interactions, suggesting that model training specialised the attention head for this purpose while other heads were more focused on the local environment. The attention weights of the second attention block are much higher at longer distances. This is consistent with focusing on high-order correlations, as features entering the second attention head are more context-enriched after passing through the first attention block.}
	\label{fig:attention_heads}
\end{figure}

Examining the attention weights of the trained band gap model for all pairs of atomic sites in the test dataset allows interrogation of the model to investigate the importance of pairwise atomic interactions at different distances (\cref{fig:attention_heads}). The attention heads of the first attention block focus on atomic sites that are close together ($<$5\,\AA{}), which is consistent with local interactions being important to material properties. The second attention head within the first block notably contains more long-range interactions, suggesting that model training specialised the attention head for this purpose while other heads were more focused on the local environment. Enforcing a cutoff limit of 5\,\AA{} on the range of the attention and retraining the model decreases performance (MAE 0.291\,eV), confirming that interactions beyond this distance meaningfully contribute to model predictions.

Meanwhile, the attention weights of the second attention block are much higher at longer distances. This is consistent with focusing on higher-order correlations, as features entering the second attention head are more context-enriched after passing through the first attention block. Notably, the model learns that the majority of significant interactions are at short range but is able to nevertheless capture significant interactions at longer distances, without having to define beforehand what constitutes a meaningful interaction. This is consistent with the decrease in performance seen when a cutoff distance is enforced. 
Enforcing a 5\,\AA{} distance cutoff limit to the attention in Site-Net decreases model performance to levels to graph-based models with the same cutoff (MAE 0.291\,eV).

While Site-Net demonstrates excellent performance on the band gap prediction task, not all tasks are expected to benefit from identical model features. Accordingly, most model features of Site-Net used in this report were deliberately chosen to be tuneable hyperparameters that can be learned (\textit{e.g.}, the learned elemental embedding), but some initial site features and interaction features were defined manually and may not be optimal. 
For example, construction of models without the Coulomb matrix in the interaction features resulted in marginal decrease in model performance on the band gap regression task, while models trained without the real-distance matrix led to reasonable training but poor test set performance. Further, the hyperparameter search was only preliminary owing to the large search space, and thus model hyperparameters used here could be far from optimal. For example, the attention weights demonstrate similar distance-dependent behaviour in attention head 1 and attention head 3 (\cref{fig:attention_heads}), so the removal of an attention head may only marginally decrease performance while significantly reducing model training time.

We have shown here that Site-Net is effective at operating on ordered crystal structures, but owing to the construction of a large supercell and removal of symmetry, the same process can also be used to examine disordered crystal structures. Disordered materials could either be treated directly (\textit{e.g.}, using the raw atom positions from a molecular dynamics simulation), or treated by constructing multiple ordered supercells (\textit{e.g.}, using Pymatgen \cite{ONG2013314}) and generating predictions for all supercell approximants. We note the predictions on the set of ordered supercells could be aggregated and subsequently interrogated using simple statistics to infer the reliability of the predictions.

\subsection{Computational Considerations}
\label{sec:computational_considerations}
The model is computationally intensive and has a quadratic dependence on the number of atoms in the crystal in terms of VRAM and computational load. The current model ingests unit cells of a maximum size of 100 atoms, and was trained using 16\,GB of VRAM, which can run on a single desktop GPU. While this was done as a proof of concept, and the model size in this report was limited by the amount of GPU VRAM, there is no fundamental limitation to the number of atoms that can be ingested beyond the amount of computational resources available to operate the larger attention blocks. 

\begin{figure}[!ht]
	\centering
	\includegraphics[width=1\textwidth]{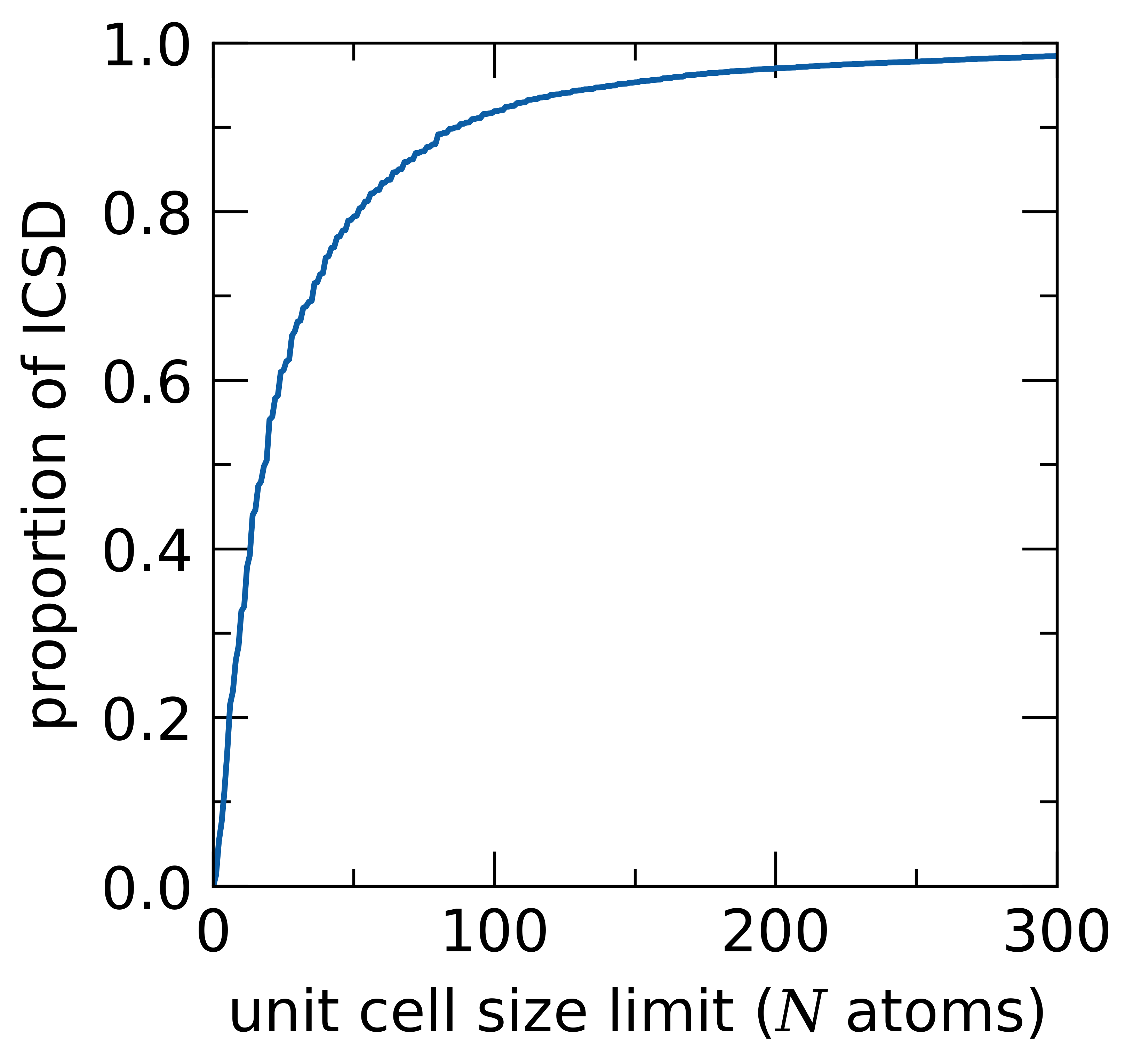}
	\caption[The proportion of crystal structures in the ICSD as a function of unit cell size limit.]{
	The proportion of all crystal structures in the ICSD as a function of unit cell size limit (\textit{i.e.}, the maximum number of atoms in the unit cell). 
	The cutoff limit of 100 atoms used in this work would allow training on 92\% of the crystal structures in the ICSD, whereas a limit of 300 atoms corresponds to 98\% of the ICSD.
	}
	\label{fig:size_limit_ICSD}
\end{figure}

100 atoms was chosen for models trained and reported here because $>$97\% of structures in the band gap dataset have unit cells with less than 100 atoms (\cref{fig:size_limit_matbench}). However, the cutoff limit of 100 atoms should also be appropriate on more general tasks using other datasets. Similar examination of $\sim$200\,000 crystal structures in the Inorganic Crystal Structure Database (ICSD) demonstrates a cutoff limit of 100 atoms would allow training on 92\% of the crystal structures in the ICSD (\cref{fig:size_limit_ICSD}).
A cutoff limit of 100 atoms provides a balance by having sufficiently large local environment for any atomic site as well as including nearly all of the data in the training set, while avoiding prohibitive batch sizes and training times. It is expected that increasing the number of atoms will increase performance up to a certain point, as the larger local environments in larger supercell models will allow the explicit encoding of increasingly long-range interactions.

There are several straightforward ways to achieve larger models. In the most simple case, larger supercells of 300 atoms could easily be treated by running on a larger GPU, for example using a high-performance computing cluster with a 128\,GB GPU. Being able to run models with a size limit of 300 atoms allows using 99.96\% of all the data for the dataset used here (\cref{fig:size_limit_matbench}), generates supercells that have a large number of atoms to encode long-range interactions (\cref{fig:atoms_in_cell}, and generates well-behaved pseudo-cubic supercells  (\cref{fig:cubic_deviation}). Straightforward changes to the architecture can also be made to achieve larger models. The first is to split the parameters of the model across multiple GPUs to increase the available VRAM and speed up training. Alternatively, parameters of the model could be offloaded from VRAM to system RAM or even high speeds solid state drives \cite{DeepSpeed}. These methodologies combined would allow a site-net implementation to scale to a local environment of arbitrary size.

More significant improvement in performance will come from better handling of the tensors in the model. Zero padding used in the current model means that the largest crystal in the dataset that determines the VRAM and computational requirements, and all crystal structures must be scaled up to the largest member. An implementation with full support for variable sized tensors will considerably reduce VRAM requirements, as it removes the need for zero-padding \cite{tensorflow2015-whitepaper}.
Specifically, for the Matbench band gap prediction task, with 100 atom supercells as used here, the VRAM requirement would be reduced by a factor of 5. This would allow either training the same model 5 times faster, or training a model that is 5 times larger (\textit{e.g.}, 500 atoms). These improvements will become increasingly large  more significant with larger supercells.

This has further benefits, because atomic sites generated in supercells could then also be treated more efficiently. Rather than calculating the attention weights explicitly on all atomic sites in the supercell, calculation of the attention weights and training could be performed for only the unique atomic sites in the initial minimal $P1$ unit cell. Specifically, in the interaction features tensor $I_{i,j,f}$, the length of $i$ would be equal to the number of atomic sites in the minimal $P1$ unit cell, and the length of $j$ would be equal to the total number of atoms in the supercell.

Finally, after Site-Net becomes more mature and the reasonable ranges of hyperparameters are outlined, the hyperparameter search space can likely be reduced, which will lead to much faster model training.

\subsection{Invariance under unit cell transformations}
\label{sec:invariance}
Every physical crystal structure can be represented using many possible unit cells or supercells. For example, the choice of unit cell setting in triclinic crystal systems is not straightforward \cite{Donnay1943}, and non-standard representations can be preferred in some circumstances (\textit{e.g.}, the use of hexagonal unit cells as opposed to primitive rhombohedral unit cells \cite{Burns201345}). Transforming between unit cells changes various parameters in the CIF, such as the atomic coordinates and unit cell parameters. If these parameters form part of the input of a machine learning model, then the choice of unit cell can lead to different predictions. This issue has gained recent attention in the literature and has prompted assessment of existing models \cite{Ropers2021, tfieldnetworks, worrall_hnets_2016}.

\begin{figure}[!ht]
	\centering
	\includegraphics[width=\textwidth]{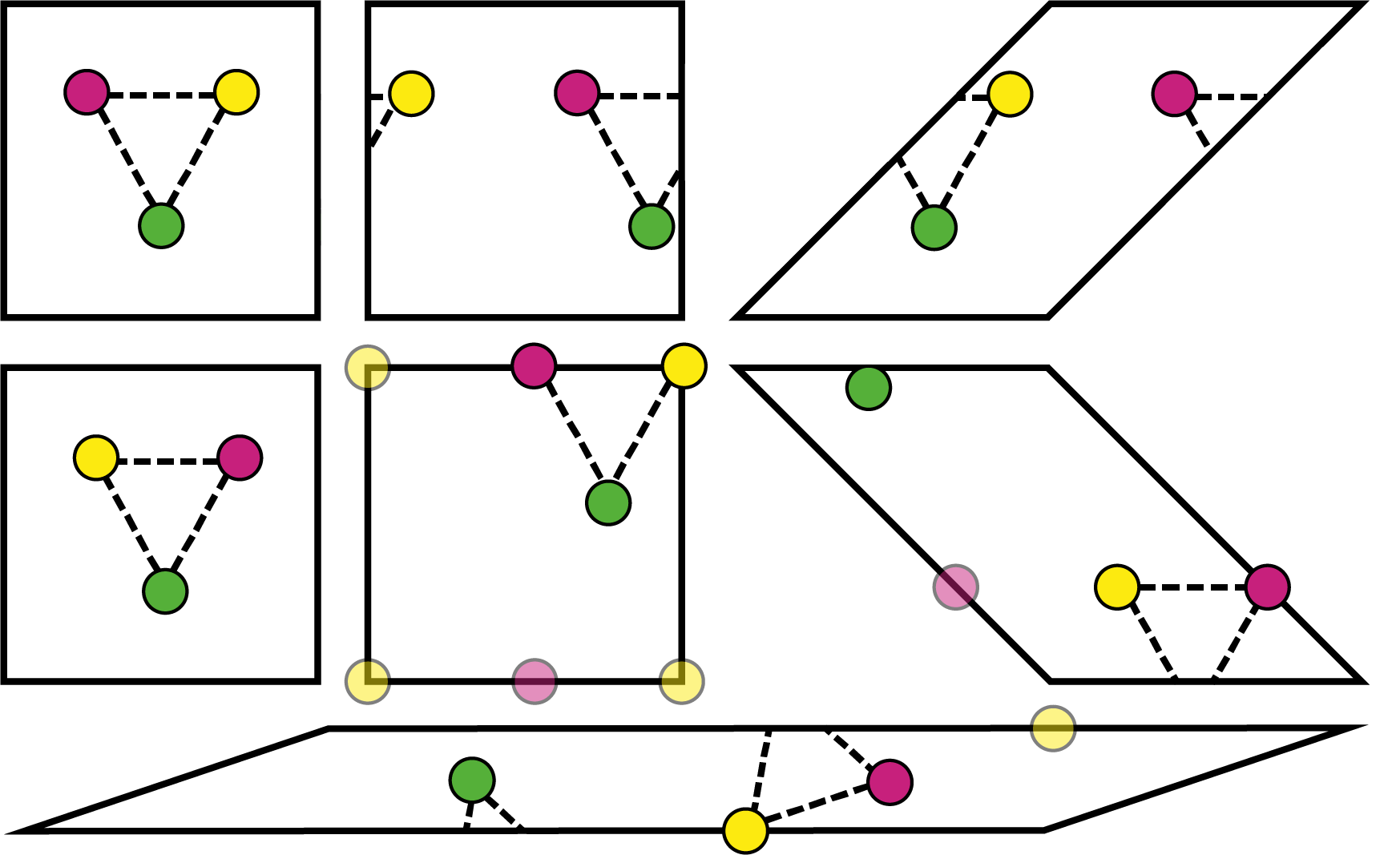}
	\vspace{-0.4cm}
	\caption[Translations, reflections, and unimodular transformations of the unit cell parameters do not change the computed interaction features.]{
    Translations, reflections, and unimodular transformations of the unit cell parameters do not change the computed interaction features. We show several transformations of a hypothetical two-dimensional unit cell containing 3 crystallographic sites. For sites on corners and edges, a position is chosen arbitrarily; equivalent choices are shown with transparency.
    Despite apparently distinct arrangements of the sites, the pairwise distances remain the same.}
	\label{fig:invariance}
\end{figure}

The design of the Site-Net model ensures that model predictions do not change under translations and unimodular transformations, described below. 
These transformations do not change the volume of the unit cell or supercell, but they nevertheless lead to unit cells that are very distinct (\cref{fig:invariance}).
Importantly, the types and quantities of crystallographic sites remain unchanged by these transformations, so while the order of crystallographic sites may change, the same set of inputs $S_{i,f}$ and $I_{i,j,f}$ will be processed by a model that is invariant to permutations.

In the case of translation, which is more straightforward, the number of crystallographic sites $S^t_{i,f}$ in the translated unit cell and their identities remain the same, but their ordering might change. Formally, it means that there is a permutation $\pi$ over the sites such that $S_{i,f}$ and $S^t_{\pi(i),f}$ are equal. Furthermore, since the distances are computed under periodic boundary conditions (\textit{i.e.}, the distance between any two sites is always the distance between an atomic site and the closest site from any self or image unit cell), the resulting interaction features $I^t_{i,j,f}$ will be identical to $I_{i,j,f}$ after the rearrangement $I^t_{\pi(i),\pi(j),f}$. Thus, the tensors $B_{i,j,f}$ and $B^{t}_{i,j,f}$, which constitute the only input to Site-Net, are identical up to a permutation. Since all operations performed in Site-Net are permutation-invariant, we arrive at the same predictions.

The same reasoning applies in the case of unimodular transformations, where we show that the number of crystallographic sites and their identities are preserved. 
A crystallographic lattice defined by the lattice vectors $V = [\vec{a}, \vec{b}, \vec{c}]$ can be generated using different sets of vectors.
A classical result from lattice theory states that multiplication of $V$ by a unimodular matrix $U$ (\textit{i.e.}, a matrix with integer coefficients and the determinant $\pm 1$) leads to vectors $V^\prime = VU$ that also generate the initial lattice \cite{Micciancio2002}. 
As the point lattice before and after transformation are identical, both unit cells (as fundamental domains) 
will have the same volume and contain a unique representative of every crystallographic site.
Therefore, the new sites $S^u_{i,f}$ of the unit cell after a unimodular transformation are identical to the original sites $S_{\pi(i),f}$ after application
of a suitable permutation $\pi$ of indices. Similarly to the case of translations, we can conclude that the the tensors $B_{i,j,f}$
and $B^{u}_{\pi(i),\pi(j),f}$ are the same, which leads to identical predictions produced by the Site-Net model.

Finally, it is important to note that Site-Net is, by design, not invariant to scale. Site-Net is designed to update its predictions by incorporating increasingly long-range interactions. Accordingly, as the supercell size is increased, attention heads will be able to examine more interactions at longer radial distances, and we expect convergence at some sufficiently long distance when all meaningful interactions are considered.

\section{Conclusions}
\label{sec:conclusions}

We present Site-Net, a transformer model for learning structure--property relationships in extended inorganic solids. Site-Net processes standard crystallographic information files
, and uses a physically motivated representation of the crystal structure as a point set of atomic sites. As many physical phenomena in extended inorganic solids arise from long-range interactions and features of the crystal structure, we build a large supercell to encode this information explicitly.
Critically, the set of atomic sites is directly ingested without any pre-defined connections, and the importance of interactions between all atomic sites is flexibly learned by the model for the prediction task presented.

The relevant structural information will differ between property prediction tasks, and the use of a custom global self-attention mechanism on all pairwise interactions of atomic sites allows Site-Net to identify important interactions and effectively deal with the all-to-all connectivity that would otherwise be overwhelming.
The attention mechanism in Site-Net works by iteratively replacing the atomic sites with context-enriched versions of themselves, which are created by aggregating the most important structural information from all other atomic sites in the crystal structure present in the supercell. 

The use of attention in Site-Net allows interrogation of the learning by examining the weights assigned to interactions at different interatomic distances. We show that for the band gap prediction task performed here, Site-Net learns from interactions that are beyond the nearest neighbour atomic sites, and that attention heads performing the attention calculations become specialized to deal with primarily short- or long-range interactions.
Further, training Site-Net where the attention has an artificial distance cutoff limit of 5\,\AA{} decreases model performance, confirming that including longer range interactions within a crystal structure meaningfully contributes to property predictions of extended inorganic materials.

We demonstrate the effectiveness of Site-Net through a band gap prediction task, as this task is heavily studied and commonly used as a benchmark for model performance. As a proof of concept, we build small supercells of 100 atoms and train Site-Net using a single consumer graphics processing unit (GPU).
Site-Net achieves a mean absolute error (MAE) of 0.251\,eV using the Matbench band gap regression dataset, and performance of the model is consistent across band gap values. Even after only a preliminary hyperparameter search and using small supercells of 100 atoms, Site-Net demonstrates competitive performance with the highest performing algorithms on the Matbench leaderboard. The performance of Site-Net is likely to improve following a more extensive hyperparameter search, and through the use of larger supercells. Both paths to improvement can be easily accommodated through changes to the way calculations are handled internally as well as through the use of larger or parallel GPUs.

Importantly, we show that explicit incorporation of long-range interactions through the use of supercells can improve the performance of machine learning models that use crystal structure to predict properties of extended inorganic solids.
Given that many physical properties result from long-range features and/or the extended nature of a crystal structure, the performance of other models on many prediction tasks may be likewise improved through similar methods, particularly where the models rely solely on the primitive unit cell.

\section{Acknowledgements}
\label{sec:others}
Work was performed using Barkla, part of the High Performance Computing facilities at the University of Liverpool, UK. The authors thank the Leverhulme Trust for funding via the Leverhulme Research Centre for Functional Materials Design. MWG thanks the Ramsay Memorial Fellowships Trust for funding through a Ramsay Trust Memorial Fellowship. 

\bibliographystyle{unsrtnat}
\bibliography{references}  

\clearpage

\beginsupplement
\title{Supporting information: Site-Net: Using global self-attention and real-space supercells to capture long-range interactions in crystal structures}
\maketitle

\section*{Construction of roughly cubic supercells with a limited number of atoms}
\label{sec:supercell_algo}

Site-net operates by recursively aggregating sites with their local environment, so it is critical that the local environment of an atomic site is well-behaved at radial distances examined by the attention heads. Accordingly, there is a soft requirement to provide each atomic site in the crystal the largest local environment possible, so a supercell is created to explicitly include longer range interactions with images of the minimal $P1$ unit cell. 

Ideally, there should be no edge effects or finite size effects owing to the construction of the supercell; the local environment of any atomic site within the minimal $P1$ unit cell should be equivalent to looking out into the infinite crystal structure. 
In practice, the size of the model is limited by computational resources, and the distance at which these edge effects begin to contribute is defined here as the self-intersection limit, which is half the shortest distance from any site to its own image in a neighbouring unit cell (or supercell).
As an example, an attention head operating on an atomic site at the centre of an orthorhombic unit cell would examine all interactions out to the edge of the unit cell, after which there are no interactions to consider for attention along that direction, resulting in a self-intersection limit equal to half the shortest unit cell parameter.
It is possible that overtraining could result from the model learning the edge effects if making use of interactions beyond this range. 

\begin{figure}[!ht]
	\centering
	\includegraphics[width=1\textwidth]{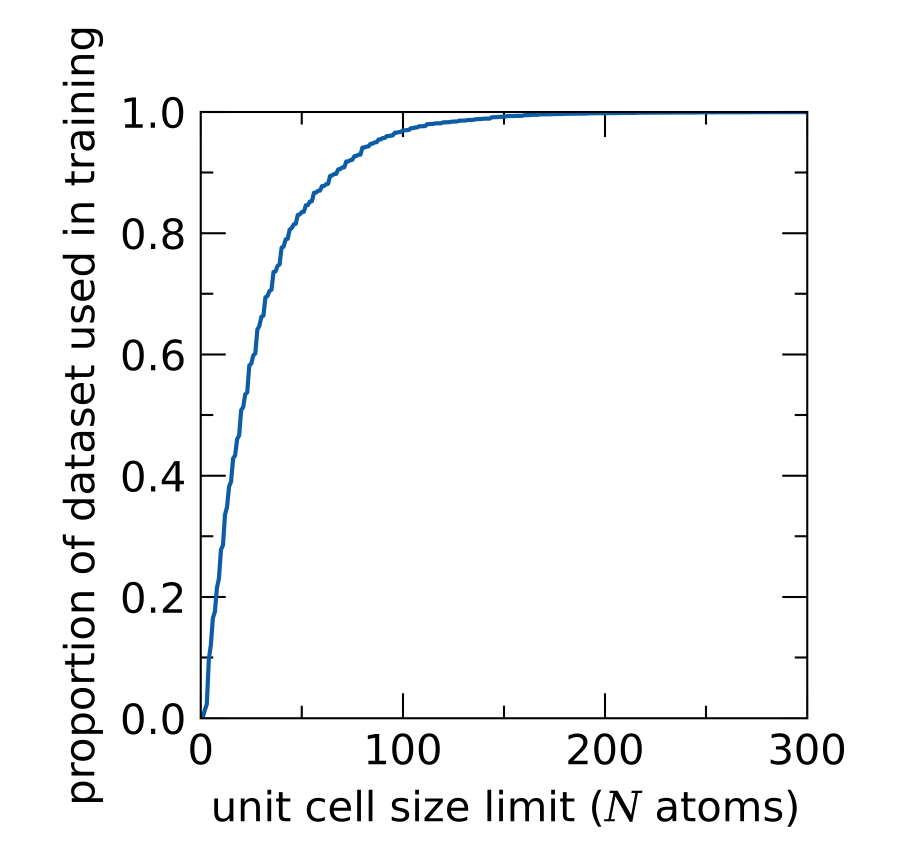}
	\caption[The proportion of crystal structures preserved in the training data of the Matbench band gap prediction task as a function of unit cell size limit.]{
	The proportion of crystal structures preserved in the training data of the Matbench band gap prediction task as a function of unit cell size limit (\textit{i.e.}, the maximum number of atoms in the unit cell). 
	Some crystals structures are lost from the training data as a function of the maximum number of atoms that are considered; the cutoff of 100 atoms used in this work allows using 97\% of the crystal structures for training. Importantly, no crystals are excluded from the test dataset when site-net is run in inference mode.
	}
	\label{fig:size_limit_matbench}
\end{figure}

\begin{figure}[!ht]
	\centering
	\includegraphics[width=1\textwidth]{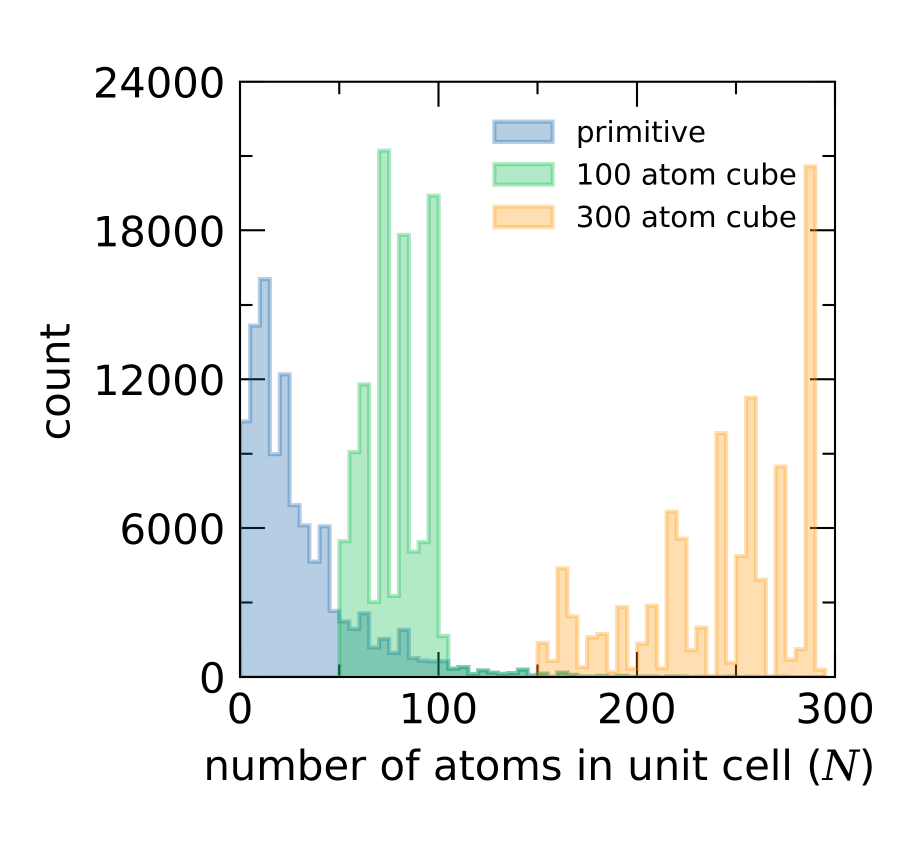}
	\caption[The number of atoms in the unit cells or supercells of crystal structures in the Matbench band gap prediction task for different unit cell size limits.]{
	The number of atoms in the unit cells or supercells of crystal structures in the Matbench band gap prediction task for different unit cell size limits (\textit{i.e.}, the maximum number of atoms in the unit cell). 
	For a given supercell size limit, the minimum number of atoms in the unit cell will be $N$/2, and the maximum will be $N$.
	}
	\label{fig:atoms_in_cell}
\end{figure}

\begin{figure}[!ht]
	\centering
	\includegraphics[width=1\textwidth]{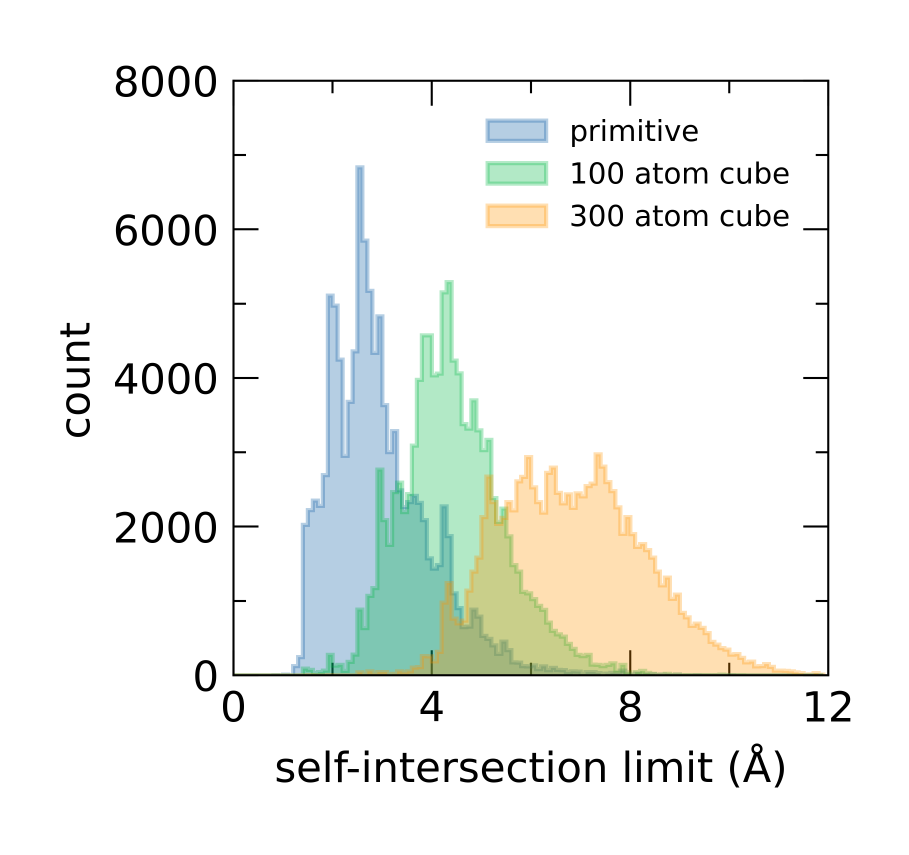}
	\caption[The self-intersection limit for unit cells or supercells of crystal structures in the Matbench band gap prediction task at different unit cell size limits.]{
	The self-intersection limit for unit cells or supercells of crystal structures in the Matbench band gap prediction task at different unit cell size limits.
	The self-intersection limit is defined here as half the minimum distance from any atomic site in the unit cell to its mirror image in a neighbouring unit cell. For orthorhombic unit cells, the self-intersection limit is half the shortest unit cell parameter.
	Radial distances longer than the self-intersection limit will lead to finite size effects, which will become more significant with distance.
	Larger supercells increase the self-intersection limit, and enable the examination of longer range interactions.
	}
	\label{fig:self-intersection}
\end{figure}

\begin{figure}[!ht]
	\centering
	\includegraphics[width=1\textwidth]{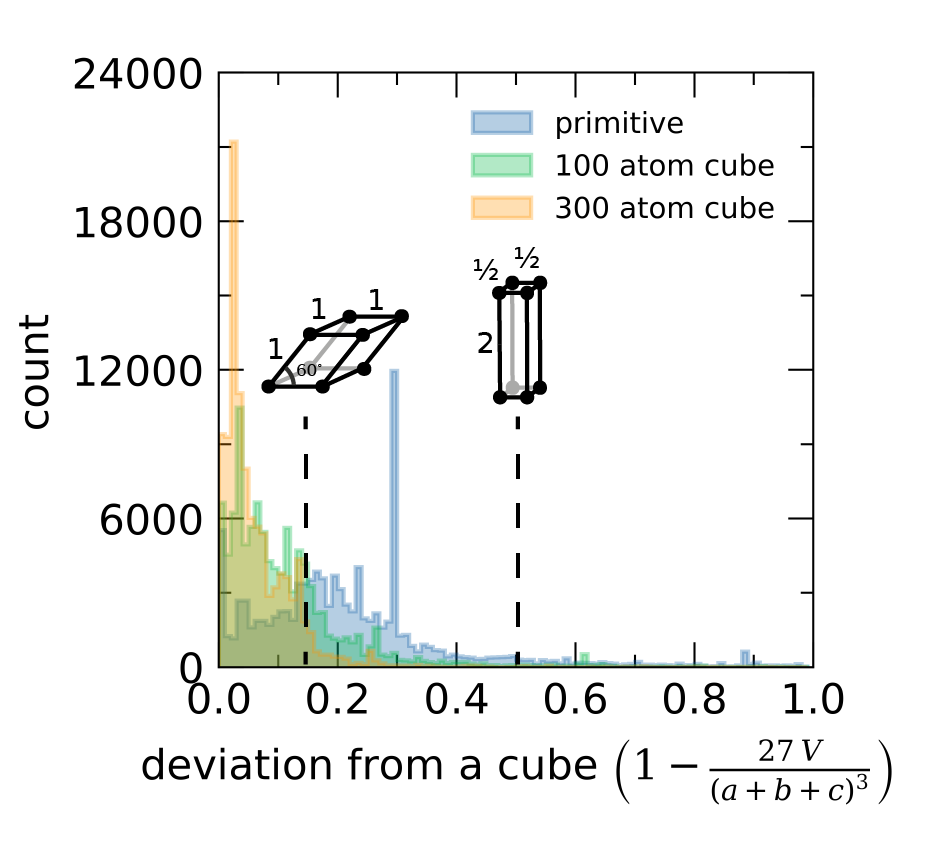}
	\caption[The deviation from a cube for unit cells or supercells of crystal structures in the Matbench band gap prediction task at different unit cell size limits.]{
	The deviation from a cube for unit cells or supercells of crystal structures in the Matbench band gap prediction task at different unit cell size limits. 
	The deviation is defined here as one minus the ratio between the volume of a unit cell and the analogous cubic unit cell with a side length equal to the average of the unit cell parameters.
	A more isotropic cell ensures the local environment of an atomic site is well-behaved at longer radial distances, and should minimize finite size effects.
	With increasing number of atoms allowed in the supercell, the distribution of cell shapes approaches a cube.
	}
	\label{fig:cubic_deviation}
\end{figure}

To maximize the self-intersection limit, all crystal structures are transformed to the largest possible supercell that is approximately cubic and contains an appropriate number of atoms -- fewer than a specified limit. We present below a simple algorithm for this task.
To explore how the supercells behave at different size limits, we build supercells of 100 atoms and 300 atoms for crystal structures within the Matbench band gap prediction dataset, and examine key features relevant to Site-Net (\cref{fig:atoms_in_cell,fig:self-intersection,fig:cubic_deviation,fig:size_limit_matbench}). In the current implementation of Site-Net, 300 atoms represents a rough maximum supercell size that could be handled by a single GPU. 
For every crystal structure, we change the maximum size limit of the supercell and then examine (a) the resulting number of atoms (\cref{fig:atoms_in_cell}), (b) the self-intersection limit at which edge-effects will begin to contribute (\cref{fig:self-intersection}), and (c) the deviation of the supercell from an ideal cube (\cref{fig:cubic_deviation}). Importantly, all features become more well-behaved with increasing supercell size, and a greater portion of the dataset becomes available for training (\cref{fig:size_limit_matbench}).

We now formally describe the transformation creating an approximately cubic supercell.
To begin, let $V^P$ be the matrix where each row is a basis vector of the minimal $P1$ unit cell. We perform Gram--Schmidt orthogonalization procedure on $V^P$ to obtain the decomposition $V^P = RQ$, where $R$ is the upper triangular matrix and $Q$ is an orthogonal matrix, \textit{i.e.}, $QQ^T = I$. Note that normalization is not performed, thus $R$ has values of 1 on the diagonal entries. Given that $R^{-1}V^P = Q$, the transformation $R^{-1}$ creates an orthogonal unit cell. This is the orthogonalization component of our transformation. 

Next, $Q$ is used to compute the scaling component of the supercell transformation, denoted as $S$. The shortest unit cell basis vector from $Q$ is iteratively incremented until any further incrementation would bring the number of atoms in the unit cell above the specified limit (100 in this work); $S$ is the diagonal matrix that performs the scaling of $Q$, and the $i$-th diagonal entry $s_{i,i}$ encodes the number of times the $i$-th lattice vector should be repeated to form the supercell.

Finally, by combining $R^{-1}$ and $S$ into $SR^{-1}$, we compute a transformation matrix that converts the minimal $P1$ unit cell $V^P$ into an approximately cubic supercell denoted $V^S$. 
Note that $R$ is invertible, as it has only values of 1 on the diagonal, so its determinant is 1, and its inverse is also an upper triangular matrix.
The combined transformation $SR^{-1}$ will have non-zero values on the diagonal; however, the matrix entries are likely to be non-integer, which is problematic because only non-singular integer matrices are guaranteed to create a valid supercell.
Accordingly, the entries of $SR^{-1}$ are rounded to the nearest non-zero integer to obtain the cubic supercell lattice parameters $C^S$, where $V^S = \round{SR^{-1}}V^P$.
This whole procedure ensures the supercell $V^S$ is valid, approximately cubic, and contains less than a specified number of atoms.

Importantly, performing the orthogonalization and scaling together is much more versatile than performing these transformations serially, as rounding matrix elements does not need to be performed until the end of the process. Effectively, performing both operations together means the minimal $P1$ unit cell $V^P$ can be tiled along directions not parallel to the unit cell basis vectors. Accordingly, the final supercell is expected to be a better approximation of a cube, which will have the largest possible self-intersection limit. 

\clearpage

\clearpage
\newgeometry{top=2cm,bottom=2cm}

\section*{Explicit architecture}

\end{document}